\documentclass[aps,prd,twocolumn,showpacs,groupedaddress]{revtex4-1}

\usepackage{amsfonts,amsmath,amssymb,mathrsfs}
\usepackage{hyperref}
\usepackage{subfigure}
\usepackage{color}
\usepackage{graphicx}  % needed for figures
\usepackage{dcolumn}   % needed for some tables
\usepackage{bm}        % for math
\usepackage[english]{babel}
\usepackage{bm}
\usepackage{latexsym}
\usepackage{dcolumn}
\usepackage{amsmath,amsfonts,amssymb}
\usepackage{graphicx,epsfig}
\usepackage{color}
\usepackage{verbatim}

\begin{document}

%%%%%%%%%%%%%%%%%%%%%%%%%%%%%%%%%%%%%%%%%%%%%%%%%%%%%%%%%%%%%%%%%%%%%%%%%%%%%%%%%%%%%%%%%%%%%
\title{Drift, Drag and Brownian motion in the Davies-Unruh bath}
%%%%%%%%%%%%%%%%%%%%%%%%%%%%%%%%%%%%%%%%%%%%%%%%%%%%%%%%%%%%%%%%%%%%%%%%%%%%%%%%%%%%%%%%%%%%%

 \author{Sanved Kolekar}
 \email{sanved@iucaa.ernet.in}
 \author{T.~Padmanabhan}
 \email{paddy@iucaa.ernet.in}
 \affiliation{IUCAA,
 Post Bag 4, Ganeshkhind, Pune - 411 007, India}

\date{\today}
%%%%%%%%%%%%%%%%%%%%%%%%%%%%%%%%%%%%%%%%%%%%%%%%%%%%%%%%%%%%%%%%%%%%%%%%%%%%%%%%%%%%%%%%%%%%%%%%%%%%%%%%
\begin{abstract}
%%%%%%%%%%%%%%%%%%%%%%%%%%%%%%%%%%%%%%%%%%%%%%%%%%%%%%%%%%%%%%%%%%%%%%%%%%%%%%%%%%%%%%%%%%%%%%%%%%%%%%%%

An interesting feature of the Davies-Unruh effect is that an uniformly accelerated observer sees an isotropic thermal spectrum of particles even though there is  a preferred direction in this context, determined by the direction of the acceleration $\textbf{g}$. We investigate the thermal \textit{fluctuations} in the Unruh bath by studying the Brownian motion of particles in the bath, especially as regards to isotropy. We find that the thermal fluctuations are anisotropic and induce different frictional drag forces on the Brownian particle depending on whether it has a drift velocity along the direction of acceleration $g$ or in a direction transverse to it. Using the fluctuation-dissipation theorem, we  argue that this anisotropy arises due to quantum correlations in the fluctuations at large correlation time scales.

%%%%%%%%%%%%%%%%%%%%%%%%%%%%%%%%%%%%%%%%%%%%%%%%%%%%%%%%%%%%%%%%%%%%%%%%%%%%%%%%%%%%%%%%%%%%%%%%%%%%%%%%
\end{abstract}
%%%%%%%%%%%%%%%%%%%%%%%%%%%%%%%%%%%%%%%%%%%%%%%%%%%%%%%%%%%%%%%%%%%%%%%%%%%%%%%%%%%%%%%%%%%%%%%%%%%%%%%%
\maketitle
\vskip 0.5 in
\noindent
\maketitle

In 1976, Unruh discovered \cite{Davies,Unruh} that a uniformly accelerated observer, moving in a flat spacetime, perceives the inertial vacuum state to be a thermal bath with a temperature $T$ related to the magnitude of its acceleration $g$ as $T=g/2\pi$. The spectrum of particles seen by the observer is known to be  isotropic in spite of the fact that the uniformly accelerated observer obviously has a preferred direction, viz. the direction of acceleration. Alternatively, one can say that the spectrum of quantum fluctuations evaluated   along the integral curves of the  Lorentz boost killing vector are isotropic. Given this result, it is  natural to ask whether small deviations in motion away from the  integral curves of the Killing vector break this isotropy. Physically, this would correspond to the following question: If one moves with, say a uniform velocity with respect to the thermal bath itself,  does the spectrum of particles as seen by the drifting observer depend on the relative orientations of the acceleration with respect to the drift velocity? [Motion relative to an isotropic spectrum will always introduce the standard Doppler anisotropy with a $\cos\theta$ dependence with respect to the direction of the velocity, even in the absence of acceleration etc. This is \textit{not} what we are interested in. We would like to know whether the spectrum depends on the relative orientation of velocity and acceleration in addition to this standard Doppler shift dependence on the direction of velocity alone.]

We will show in this paper that the spectrum of particles seen by the drifting observer \textit{does} indeed depend on whether she is moving along $\hat{\mathbf{g}}$ or in a direction transverse to it. In particular, the spectrum in the two cases match with each other only in the high frequency regime while they differ significantly at low frequencies. 

These anisotropies also reflect in the Brownian motion of a particle in the Davies-Unruh thermal bath which can be seen as follows:
The Brownian motion of a particle in a thermal bath is described by the Langevin equation 
\begin{eqnarray}
 \frac{dv}{dt} = - \int^{\ t}_{-\infty} \gamma(t-t^\prime) v(t^\prime) dt^\prime + {\cal F}(t) + F_{ext}
\label{langevineqn}
\end{eqnarray}
where $\gamma(t-t^\prime)$ is the dissipation function, ${\cal F}(t)$ is the random part of the force per unit mass due the thermal fluctuations and $F_{ext}$ is the external force per unit mass acting on the particle. We calculate the spectrum of particles seen by a drifting observer in the Davies-Unruh thermal bath and show that the observer experiences a drag force of the form $F = -\gamma v$ opposite to its direction of drift. This allows us to read-off $\gamma(\omega)$ which is the spectral decomposition of the dissipation function $\gamma(t-t^\prime)$ appearing in Eq.(\ref{langevineqn}) obtained by a Fourier transform. Then the result derived below implies that $\gamma_{\hat{\mathbf{g}}}(t-t^\prime) \neq \gamma_{\hat{\mathbf{g}_{\perp}}}(t-t^\prime)$ at large correlation time scales. This anisotropy is purely quantum mechanical in origin. Consequences of this effect on Brownian motion are discussed in the last section.

We will now proceed to describe the setup we use for the calculation. We consider a two level quantum system linearly coupled to a quantum field (known as the Unruh-Dewitt detector \cite{unruhdet}) and analyze its excitation probability when it is in motion. It is well known that using linear perturbation theory the angular dependence of the  excitation probability can be expressed in the form of an integral given by \cite{Kolbenstvedt} 
\begin{eqnarray}
\frac{d R(E)}{d \Omega} &=&\int \int  \frac{d \tau d\tau^{\prime} \ e^{-i E (\tau - \tau^{\prime})}}{\left[ (T_\tau-T_{\tau^\prime}) - \hat{\mathbf{n}} \cdot ({\bf X}_{\tau} -{\bf X}_{\tau^\prime}) - i \epsilon \right]^2}
\label{angularresponse}
\end{eqnarray}
where $T_\tau = T(\tau)$ and ${\bf X}_{\tau} = {\bf X}(\tau)$ are evaluated on the trajectory of the detector. The angular frequency response, $dR(E)/d\Omega$, determines the number density of particles emitted by the detector in a given solid angle $d\Omega$. It has been  shown that this direction sensitive detector correctly reproduces the isotropic nature of the Unruh bath when viewed from the rest frame of the bath \cite{Kolbenstvedt} and hence is ideally suited for our present purpose. We next choose a suitable trajectory for the detector such that it matches with the mean trajectory of the Brownian particle in a thermal bath and calculate the angular response for the detector to obtain $\gamma(E)$. (Note that since $\gamma(t - t^\prime)$ is an intrinsic property of the thermal bath, it is independent of the detailed model of the Brownian particle chosen.) We will now proceed to calculate $\gamma(t - t^\prime)$ first for the case of a detector drifting parallel to the direction of acceleration $\textbf{g}$ of the Rindler observer. 
%%%%%%%%%%%%%%%%%%%%%%%%%%%%%%%%%%%%%%%%%%%%%%%%%%%%%%%%%%%%%%%%%%%%%%%%%%%%%%%%%%%%%%%%%%%%%%%%%%%%%%%%
\subsection{Detector with a drift parallel to acceleration}
%%%%%%%%%%%%%%%%%%%%%%%%%%%%%%%%%%%%%%%%%%%%%%%%%%%%%%%%%%%%%%%%%%%%%%%%%%%%%%%%%%%%%%%%%%%%%%%%%%%%%%%%

To study the Brownian motion of a particle in a thermal bath with a drift along the direction of acceleration, we choose a trajectory such that the Rindler observer, who is in the rest frame of the bath, perceives this detector as moving with a relative velocity $v$ in the direction of acceleration, say the ${\bf \hat{x}}$ -direction. The trajectory \textit{in the Rindler frame} is then
\begin{eqnarray}
x = v \gamma t^\prime  \; , \; \; \;\;\;\;\;  t = \gamma t^\prime
\label{yx-traj-rind}
\end{eqnarray}
where $v$ is a constant and defines the the relative velocity of the Brownian particle with respect to the rest frame of the Davies-Unruh bath. A suitable coordinate transformation between the static Rindler observer and the drifting observer is 
$x=\gamma(x^\prime+v t^\prime)$ and $t = \gamma ( t^\prime + v x^\prime)$. In these coordinates, the trajectory is simply $x^\prime = 0$ and the metric becomes
\begin{eqnarray}
 ds^2 &=&e^{2g\gamma x^\prime} e^{2g\gamma vt^\prime} \left[ -dt^{\prime 2} + dx^{\prime 2} \right] + dy^2 + dz^2
\label{yx-metric2}
\end{eqnarray}
From the metric, it is clear that the time co-ordinate $t^{\prime}$  is \textit{not} the proper time $\tau$ of the drifting observer but is related to the proper time $\tau$ through $\tau = -(1 - e^{g \gamma v t^{\prime}})/(g\gamma v)$. For our purpose, it is enough to  consider drift velocities  which are very small in magnitude and we will work in the regime of linear $v$ assuming $v \ll 1$. Before proceeding further let us first quantify  how small a $v$  we need to consider. Let us assume that the detector coupling to be switched-on during the time interval $(-L,L)$ where $L$ is a very large but finite number such that $EL \gg 1$ holds. This condition basically ensures that the contour integration over $t-t^\prime$ in Eq.(\ref{angularresponse}) vanishes over semi-circle in the lower half of the complex $u$ plane. We then choose the velocity parameter $v$ such that $gvL \ll 1$. These conditions lead to $v \ll E/g$ which tell us that the low frequencies (i.e., low compared to $g$) probed by the detector, in this approximation, are limited by choice of the parameter $v$. However, we can still choose a non-zero $v$ such that $v \ll E/g \ll 1$ which allows us to probe the effects in the low frequency regime. Under this approximation, the proper time $\tau$ to linear order in $v$ is
\begin{equation}
\tau = t^\prime + \frac{gv}{2}t^{\prime 2}
\label{yx-propertime}
\end{equation}
Using the above relation, the magnitude of the proper acceleration for the trajectory in Eq.(\ref{yx-traj-rind}) can be found, upto linear order in $v$, to be 
\begin{equation}
g(\tau) = \sqrt{a^i a_i} = g(1 + g v \tau)
\label{yx-acc}
\end{equation}
The time dependence of acceleration can be understood by noticing that the observer following the trajectory given by Eq.(\ref{yx-traj-rind}) crosses different $x =$ constant surfaces at different times. This corresponds to crossing different uniformly accelerated hyperbolic trajectories at different times, which is essentially described by Eq.(\ref{yx-acc}). Further, the metric in Eq.(\ref{yx-metric2}) can now be expressed in terms of the proper time and we get
\begin{eqnarray}
 ds^2 &=&e^{2g\gamma x^\prime} \left[ -d\tau^{2} + (1+gv\tau) dx^{\prime 2} \right] + dy^2 + dz^2
\label{yx-metric}
\end{eqnarray}
(It is interesting to note that metric of the 2-dimensional $x^\prime-\tau$ hypersurface is conformal to a 2-dimensional FRW metric of an expanding universe with scale factor equal to $g(\tau)/g$.) We will later show that the factor $g(\tau)/g$ acts as the blue-shifting (or red-shifting depending on the sign of $v$)factor for the temperature as seen by the drifting detector. This blueshift in temperature accounts for the time dependent particle production which we expect due to the non-static nature of the above metric.  

We will now proceed to calculate the angular response for the detector using the expression in Eq.(\ref{angularresponse}). In Eq.(\ref{angularresponse}), the integration variables are in terms of the proper time $\tau$. However, it turns out to be easier (algebraically) to perform the contour integral using the time co-ordinate $t^\prime$ in Eq.(\ref{yx-metric2}). We use Eq.(\ref{yx-propertime}) to make a change of variable in Eq.(\ref{angularresponse}) and obtain:
\begin{eqnarray}
 \frac{d R(E)}{d \Omega} &=& \int \int \frac{ds \ du \; \left(1 + 2gvs \right) e^{-i Eu \left[1 + gvs  \right]}}{\left[ T-T^\prime - \cos{\theta}\left(X - X^\prime \right) - i \epsilon \right]^2}
\label{yx-response2}
\end{eqnarray}
where we have defined $u = (t_1^\prime - t_2^\prime)$ and $2s = (t_1^\prime + t_2^\prime)$. To evaluate the above integral we write the trajectory given in Eq.(\ref{yx-traj-rind}) in terms of the inertial coordinates expressed as function of $t^\prime$ as
\begin{eqnarray}
T(t^\prime) =\frac{(1+gvt^\prime)}{g}\sinh{g t^\prime} \; ; \;\;\;\;  X = \frac{(1+gvt^\prime)}{g}\cosh{g t^\prime} \nonumber
\end{eqnarray}
Substituting the above expressions in Eq.(\ref{yx-response2}) and expanding the integrand upto linear order in $v$ results in an integral which involves closing the contour in the lower half complex $u$ plane and evaluating the residues at the poles of the integrand. We assume that the linear order result can be obtained by first expanding the integrand to order $v$  and then evaluating the residues. The response rate of the detector is then given by
\begin{eqnarray}
 \frac{d \dot{R}(E)}{d \Omega} &=& \int\frac{(g^2/4)F^{2}(\theta,s) e^{-i \bar{E} u}du}{\left[ \sinh{(\frac{gu}{2})} - i \epsilon \right]^2} +  \frac{d \dot{R_v}(E)}{d \Omega} %\nonumber \\
\label{yx-res-total}
\end{eqnarray}
where 
\begin{eqnarray}
 \frac{d \dot{R_v}(E)}{d \Omega} &=& \int\frac{(g^2/4) e^{-i \bar{E} u} \ vu  \cosh{\left(\frac{gu}{2}\right)} du}{F^{-3}(\theta,s) \bar{F}(\theta,s) \left[\sinh{(\frac{gu}{2})} - i \epsilon \right]^3}
\end{eqnarray}
and
\begin{eqnarray}
F(\theta,s) &=& \left[\cosh{gs} - \cos{\theta} \sinh{gs} \right]^{-1} \nonumber \\
\bar{F}(\theta,s) &=& \left[\cos{\theta} \cosh{gs} - \sinh{gs} \right]^{-1}
\label{yx-ftheta}
\end{eqnarray}
Here, $\bar{E} = E\left[1 + gvs  \right] $ and the overhead ($^{.}$) denotes differentiation w.r.t $s$. The first term in Eq.(\ref{yx-res-total}) in the Taylor expansion is the familiar term one usually gets while calculating the response of a detector moving on the Rindler trajectory, except that we get $\bar{E}$ instead of $E$ in the exponential. Evaluating the integral in the first term then leads to the Planckian distribution $E/(\exp{\beta(s) E} - 1)$ of particles with $\beta^{-1}(s) = g(s)/2\pi$. The time dependence of the temperature is easily understood in this case as being due to the gravitational redshift when the detector crosses different $x =$ constant surfaces at different times. As is well known, the thermal spectrum of particles seen by the Rindler observer is isotropic, hence we expect the first term also to be isotropic since it differs from the former case only in the time dependence of temperature $\beta^{-1}$. This will be evident from our analysis below. 

The second term, which is linear in the velocity, turns out to be anisotropic in the spectrum and leads to the drag force on the Brownian particle. To calculate the required response function, we begin by first performing the contour integral over $u$. Due to the Taylor expansion, this turns out to be a simple procedure of evaluating the residue of the integrand at the same poles one gets, but now of one order higher, while evaluating the first term in Eq.(\ref{yx-res-total}). The response function is then obtained to be
\begin{eqnarray}
 \frac{d \dot{R_v}(E)}{d \Omega} &=& \frac{3}{4 \pi} \frac{\gamma_x(E)}{E} v  F^{3}(\theta,s)\bar{F}^{-1}(\theta,s)
\label{yx-res2}
\end{eqnarray}
where,
\begin{eqnarray}
\gamma_x(E) &=& \left(\frac{4}{3}\right) \frac{\pi^2 \beta E^3 e^{-\beta E}}{\left( 1 - e^{-\beta E}\right)^2} \left[ 1 - \frac{2}{\beta E} + \frac{2e^{-\beta E}}{\beta E} + \frac{2\pi^2}{\beta^2 E^2} \right] \nonumber \\
 \label{yx-dissipation}
\end{eqnarray}
Hereafter, for notational simplicity, we suppress the time-dependence in $\beta (s)$. We will show below that $\gamma_x(E)$ is the spectral decomposition of the dissipation function discussed after Eq.(\ref{langevineqn}). Note that the angular dependence of the response function in above Eq.(\ref{yx-res2}) is in terms of the solid angle $d \Omega$ defined in the inertial frame. One would have expected the  angular response to be time-independent (except for the time dependence in $\beta(s)$) whereas from the $F(\theta,s)$ dependence, it appears as though it depends on the proper time of the trajectory in a complicated manner. But recall that the natural frame of reference to consider when studying Brownian motion would be the rest frame of the thermal bath itself which, in the present case, is the Rindler frame. As we will show, all the peculiar behaviour goes away, when we view it in the Rindler observer's frame.  The relevant transformation between the angular co-ordinates of the Minkowski frame $(\theta, \phi)$ and the Rindler frame $(\theta_R, \phi_R)$ is given by
\begin{eqnarray}
 d\Omega_R &=& \frac{d\Omega}{\gamma_R^2 (1 - v_R \cos{\theta})^2} ; \;\; \sin{\theta_R} = \frac{\sin{\theta}}{\gamma_R (1 - v_R \cos{\theta})} \nonumber \\
\cos{\theta_R} &=& \frac{\cos{\theta} - v_R}{(1 - v_R \cos{\theta})}
\label{angle-trans}
\end{eqnarray}
where 
$v_R = \tanh{gs}$ is the relative co-ordinate velocity of the Rindler observer with respect to the inertial observer and $\gamma_R = 1/(1-v_R^2)$. From Eq.(\ref{yx-ftheta}), we can note that $F(\theta,s) = \gamma_R (1 - v_R \cos{\theta})$ and $\bar{F}(\theta,s) = \gamma_R (\cos{\theta} - v_R)$ and hence we get,
\begin{eqnarray}
 \frac{d \dot{R_v}(E)}{d \Omega_R} &=& \frac{3}{4\pi} \frac{\gamma_x(E)}{E} v \cos{\theta_R}
\label{yx-res}
\end{eqnarray}
The right hand side of the above equation tells us the number of particles of energy $E$ hitting the detector per unit time in a given direction $\hat{\mathbf{n}}(\theta_R,\phi_R)$. The force $\textbf{F}_{\hat{\mathbf{n}}}$ from the $-\hat{n}$ direction on the detector due to this bombardment is then $E (d\dot{R_v}(E)/d \Omega_R)$. So, the net force on the Brownian particle will be:
\begin{eqnarray}
 F_x &=& - \gamma_x(E) v 
\label{yd-res}
\end{eqnarray}
Hence, $\gamma_x(E)$ defined in Eq.(\ref{yx-dissipation}) is the spectral decomposition of the required  dissipation function $\gamma_x(t-t^\prime)$. 

We will next calculate $\gamma(t - t^\prime)$  for a detector drifting transverse to the direction of acceleration $g$ of the Rindler observer.
 
%%%%%%%%%%%%%%%%%%%%%%%%%%%%%%%%%%%%%%%%%%%%%%%%%%%%%%%%%%%%%%%%%%%%%%%%%%%%%%%%%%%%%%%%%%%%%%%%%%%%%%%%
\subsection{Detector with a drift transverse to the acceleration}
%%%%%%%%%%%%%%%%%%%%%%%%%%%%%%%%%%%%%%%%%%%%%%%%%%%%%%%%%%%%%%%%%%%%%%%%%%%%%%%%%%%%%%%%%%%%%%%%%%%%%%%%

To study this case, we choose a trajectory such that the Rindler observer will perceive the detector as moving with a relative velocity $v$ in the direction transverse to the acceleration, say the ${\bf \hat{y}}$ - direction. The trajectory in the Rindler frame is then
\begin{eqnarray}
x = 0  \; , \; \; \;\;\;\;\;  t = \gamma \tau \; , \; \; \;\;\;\;\;  y = v \gamma \tau
\label{yd-traj-rind}
\end{eqnarray}
where the parameter $\tau$ is the proper time of the observer moving on the above trajectory.

To calculate the angular response for the detector drifting along the $y$ axis we proceed in a manner similar to the case of the $x-$ drift. However, as we will see, the angular response for the $y-$ drift is relatively easier to calculate due to the fact that the $t^\prime$ co-ordinate is the proper time parameter in this case. We begin by expressing the trajectory Eq.(\ref{yd-traj-rind}) in the inertial coordinates as  
\begin{eqnarray}
T=\frac{1}{g}\sinh{g\gamma \tau}  \; ;\;\;\;  X = \frac{1}{g}\cosh{g\gamma \tau}\; ;\; \;\;  Y = v \gamma \tau
\end{eqnarray}
Substituting in Eq.(\ref{angularresponse}) and  expanding upto linear order in $v$, we get
\begin{eqnarray}
 \frac{d \dot{R}(E)}{d \Omega} &=& \int\frac{(g^2/4)F^{2}(\theta,s) e^{-i E u}du}{\left[\sinh{(\frac{gu}{2})} - i \epsilon \right]^2} +  \frac{d \dot{R_v}(E)}{d \Omega} %\nonumber \\
\label{yres-total}
\end{eqnarray}
where 
\begin{eqnarray}
 \frac{d \dot{R_v}(E)}{d \Omega} &=& \int\frac{e^{-i E u} v u \sin{\theta}\cos{\phi} du}{F^{-3}(\theta,s) \left[ \sinh{(\frac{gu}{2})} - i \epsilon \right]^3}
\end{eqnarray}
As in the $x$ drift case, the velocity independent term in Eq.(\ref{yres-total}) gives the usual isotropic Planckian distribution $E/(\exp{\beta E} - 1)$ of particles as seen by a detector moving on the Rindler trajectory and will not lead to a drag force on the detector. The second term linear in the velocity is anisotropic and leads to the required drag force. Performing the contour integral over $u$, the response function then becomes
\begin{eqnarray}
 \frac{d \dot{R_v}(E)}{d \Omega} &=& \frac{3}{4\pi} \frac{\gamma_y(E)}{E} \left(v \sin{\theta}\cos{\phi} \right)  F^3(\theta,s)
\label{yd-res2}
\end{eqnarray}
where,
\begin{eqnarray}
 \gamma_y(E) &=& \frac{4\pi^2 \beta E^3 e^{-\beta E}}{3\left( 1+ e^{-\beta E}\right)^2} \left[ 1 - \frac{2}{\beta E} - \frac{2e^{-\beta E}}{\beta E} + \frac{3\pi^2}{\beta^2 E^2} \right] \label{yd-dissipation} \nonumber \\
\end{eqnarray}
(The angular response for the detector drifting along the $y-$ axis has been obtained before in \cite{russotownsend} in the context of an accelerating braneworld model.) We will now proceed to calculate the drag force and the dissipation function and compare it with the one we obtained in the case of the $x-$ drift. We first write the response function in terms of the angular co-ordinates of the Rindler observer using Eq.(\ref{angle-trans}) and Eq.(\ref{yx-ftheta}). We get:
\begin{eqnarray}
 \frac{d \dot{R_v}(E)}{d \Omega_R} &=& \frac{3 \pi}{2} \frac{\gamma_y(E)}{E^3} \left(v \sin{\theta_R}\cos{\phi_R} \right)
\label{yd-res}
\end{eqnarray}
Hence, the net drag force on the detector in the $y$ direction is
\begin{eqnarray}
 F_y &=& - \gamma_y(E) v 
\label{yd-res}
\end{eqnarray}
 Hence, $\gamma_y(E)$ defined in Eq.(\ref{yd-dissipation}) is the spectral decomposition of the required  dissipation function $\gamma_y(t-t^\prime)$.

%%%%%%%%%%%%%%%%%%%%%%%%%%%%%%%%%%%%%%%%%%%%%%%%%%%%%%%%%%%%%%%%%%%%%%%%%%%%%%%%%%%%%%%%%%%%%%%%%%%%%%%%
\subsection{Discussion and Conclusions}
%%%%%%%%%%%%%%%%%%%%%%%%%%%%%%%%%%%%%%%%%%%%%%%%%%%%%%%%%%%%%%%%%%%%%%%%%%%%%%%%%%%%%%%%%%%%%%%%%%%%%%%%

The derivation presented in this letter provides a calculation of the dissipation function $\gamma(t-t^\prime)$ for the Davies-Unruh bath from first principles. Surprisingly, one finds that $\gamma(t-t^\prime)$ is anisotropic and depends whether one is moving in the direction of the acceleration or transverse to it. The Davies-Unruh effect itself arises due to the difference in the pattern of vacuum fluctuations seen by inertial and accelerated observers. Therefore we can interpret this anisotropy as due to the dependence of the fluctuation pattern on both velocity and acceleration which will depend on the relative orientation between the two.
This interpretation is strengthened by the fact that Eq.(\ref{yx-dissipation}) and Eq.(\ref{yd-dissipation}), for the dissipation function $\gamma_x$ and $\gamma_y$ respectively, agree in the high frequency regime in which $\beta E \gg 1$.  This feature is expected (and a cross-check on the calculation) because in the high frequency regime the quantum 
correlations will become more and more local and the difference in orientation will have less effect. 

In fact, no anisotropy of the type we are discussing will arise if we analyze the Brownian drift of a particle in a box containing massless particles in thermal equilibrium in a constant gravitational field, say, along the ${\bf \hat{x}}$ direction. In this case the equilibrium temperature of the gas will vary along $x$ axis in accordance with the Tolman \cite{Tolman} relation. An observer, drifting along the direction transverse to $\hat{\mathbf{g}}$,  will see the spectrum  to be Doppler shifted as a consequence of the standard Doppler shift in frequencies \cite{peebles} and will not experience the variation of temperature due to Tolman factor. 
An observer drifting along the direction of $\hat{\mathbf{g}}$ will experience, in addition to this standard Doppler shift, an extra gravitational redshift in the frequencies due to her motion along the gravitational field. However, the gravitational redshift factor will only make the temperature $\beta^{-1}(x(t))$ (obeying the Tolman relation) to be time dependent  and a function of the position of the observer along $\hat{\mathbf{x}}$ which is similar to the time dependence we found in the case of parallel drift in the Davies-Unruh bath. It is easy to see that the \textit{form of the frequency response} as seen by both the drifting observers would be the same  and isotropic in the sense that it depends only on the direction of $\hat{\mathbf{v}}$ and not on its relative orientation with respect to $\hat{\textbf{g}}$. (Of course, the spectrum will vary with a $\cos\theta$ dependence with respect to the direction of the velocity in both cases  but this is not the anisotropy we are interested in.) This is because such an analysis is essentially `classical' in the sense that it can be reproduced by assuming the existence of a bunch of particles in a region with a particular classical distribution function and doing just the kinematics. No quantum correlations come into play. Clearly our result agrees with the above analysis and the anisotropy vanishes in this limit.

As one moves away from the high frequency end of the spectrum, the responses of the two detectors start to differ significantly. When $\beta E \ll 1$, we have $\gamma_x \sim \beta^{-2} e^{-\beta E}/(\beta E)$ while $\gamma_y \sim \beta^{-2} (\beta E) e^{-\beta E}$. Thus the energy dissipated by a Brownian particle due to friction is greater by a fraction $1/(\beta E)^2$ because of its mean velocity in the $x$ direction than the corresponding energy dissipated in the $y$ direction.

Let us consider this result from the perspective of the fluctuation-dissipation theorem \cite{kubo} which relates the dissipation function to the two point correlation function of the random force ${\cal F}$ at different times as
\begin{eqnarray}
\gamma(\omega) &=& \frac{1}{2T}\int^{\infty}_{-\infty} \langle {\cal F}(t){\cal F}(t+t^\prime)\rangle e^{-i\omega t^\prime} dt^\prime 
\label{fluctuation-dissipation}
\end{eqnarray}
In the present context, since we have determined the dissipation function from first principles, we can now determine the characteristics of the fluctuation (or the noise) $K(u) = \langle {\cal F}(t){\cal F}(t+u)\rangle$ intrinsic to the system using the above equation. We find that, at small $((u/\beta) \rightarrow 0)$ correlation times --- which corresponds to high frequency range --- we get $K_x\approx K_y$. On the other hand, for very long $(u)$ correlation times --- which corresponds to the low frequency range --- both $K_x$ and $K_y$ vanishes with $K_x(u)/K_y(u)\propto u^2$ because of the different $E$ scaling in $\gamma_x$ and $\gamma_y$.
The behaviour $K(u) \rightarrow 0$ for $u \rightarrow \infty$ signifies that a Brownian particle which is in equilibrium with the thermal bath loses the memory of its path history (or its initial conditions) as time progresses. But our calculations also show that even though the correlations do vanish at large correlation times, there are still quantum correlations present in the case of $x$ drift which go towards zero at a slower rate as compared to the $y$ drift.

We know that the Unruh effect is closely related to the Hawking radiation of a non-extremal black hole horizon, cosmological de-Sitter horizon, etc. Hence, we would expect similar anisotropies to be present in the thermal fluctuations in the thermal bath near these horizons too. Particularly, the trajectories of particles near the horizon would be affected differently for different orientations of acceleration and velocity leading to anisotropic dissipation effects. It will be interesting to investigate consequences of this feature.

%%%%%%%%%%%%%%%%%%%%%%%%%%%%%%%%%%%%%%%%%%%%%%%%%%%%%%%%%%%%%%%%%%%%%%%%%%%%%%%%%%%%%%%%%%%%%%%%%%%%%%%
\subsection{Acknowledgments}
%%%%%%%%%%%%%%%%%%%%%%%%%%%%%%%%%%%%%%%%%%%%%%%%%%%%%%%%%%%%%%%%%%%%%%%%%%%%%%%%%%%%%%%%%%%%%%%%%%%%%%%
SK is supported by a Fellowship from the Council of Scientific and Industrial Research (CSIR), India. TP's research is partially supported by J.C.Bose Research grant.
       
%%%%%%%%%%%%%%%%%%%%%%%%%%%%%%%%%%%%%%%%%%%%%%%%%%%%%%%%%%%%%%%%%%%%%%%%%%%%%%%%%%%%%%%%%%%%%%%%%%%%%%%                                                                                    

\end{document}